\documentclass[12pt]{article}%
\usepackage{amsmath}
\usepackage{amsfonts}
\usepackage{amssymb}
\usepackage{graphicx}
\usepackage{float}%
\begin{document}

\title{Modulated phases in a spin model with Dzyaloshinskii-Moriya interactions}
\author{William de Castilho* and S. R. Salinas\\Instituto de F\'{\i}sica, Universidade de S\~{a}o Paulo, \\S\~{a}o Paulo, SP, Brazil}
\maketitle

\begin{abstract}
We analyze the phase diagram of an elementary statistical lattice model of
classical, discrete, spin variables, with nearest-neighbor ferromagnetic
isotropic interactions in competition with chiral interactions along an axis.
At the mean-field level, we show the existence of paramagnetic lines of
transition to a region of modulated (helimagnetic) structures. We then turn to
the analysis of the analogous problem on a Cayley tree. Taking into account
the simplicity introduced by the infinite-coordination limit of the tree, we
explore several details of the phase diagrams in terms of temperature and a
parameter of competition. In particular, we characterize sequences of
modulated (helical) structures associated with devil's staircases of a fractal character.

\let\thefootnote\relax\footnotetext{*william2.castilho@usp.br}


\end{abstract}

\section{Introduction}

Competing interactions are known to lead to very rich phase diagrams, with
critical and multicritical points and sequences of spatially modulated
structures \cite{Seul1995}\cite{Andelman2009}. The ANNNI model, which stands
for axial-next-nearest neighbor Ising model, is perhaps the best investigated
statistical lattice model with competing interactions \cite{Bak82}%
\cite{Selke88}\cite{Yeomans88}. In the standard ANNNI model, there are
ferrromagnetic interactions between pairs of nearest-neighbor Ising spin
variables on the sites of a cubic lattice, with the addition of competing
antiferromagnetic interactions between second-neighbor spin pairs along an
axial direction. In terms of temperature and a parameter to gauge the
competition, the ANNNI model exhibits perhaps the richest phase diagram of the
literature, with a wealth of modulated phases and the onset of fractal
structures, which have been called devil's staircases \cite{Selke1992}%
\cite{Nascimento2014}.

In the present work we were motivated by recent interest in different forms of
helimagnetism \cite{Rossler2010}\cite{Nagaosa2013}, which can be explained in
terms of a chiral mechanism proposed by Dzyaloshinskii and Moriya (DM) a long
time ago \cite{Izyumov1984}. According to this DM mechanism, interactions are
restricted to first neighbors, but with vector spin variables on each lattice
site. Modulation effects come from the competition between the usual
Heisenberg exchange, of scalar nature, and an additional exchange associated
with the vector products of the DM interactions. Different formulations of
spin models including competing exchange and DM interactions have indeed been
used to account for the monoaxial helical patterns of magnetic compounds
\cite{Kishine2005}\cite{Shinozaki2016}. We were also motivated by early work
for special clock models with chiral interactions, which have been
investigated by different techniques \cite{Huse1981}\cite{Ostlund1981}%
\cite{Ottinger1983}. In contrast to the ANNNI model, elementary lattice model
systems associated with the DM mechanism have not been subjected to more
detailed statistical mechanics calculations. We then believe that there is
still room to analyze some simple and schematic version of these spin models
with DM interactions, and to show that they are capable of displaying a phase
diagram with many sequences of modulated structures.

We consider a ferromagnetic version of a classical Heisenberg spin model with
DM interactions, in zero external field, and restrict the competition to a
single lattice axis. The spin Hamiltonian of this simple monoaxial lattice
system is written as
\begin{equation}
\mathcal{H}_{DM}=-J_{0}\sum_{\left(  \overrightarrow{r},\overrightarrow
{r}^{\prime}\right)  }^{\left(  x,y\right)  }\overrightarrow{S}%
_{\overrightarrow{r}}\cdot\overrightarrow{S}_{\overrightarrow{r}^{\prime}%
}-J\sum_{\left(  \overrightarrow{r},\overrightarrow{r}^{\prime}\right)
}^{\left(  z\right)  }S_{\overrightarrow{r}}^{z}\,S_{\overrightarrow
{r}^{\prime}}^{z}-D\sum_{\left(  \overrightarrow{r},\overrightarrow{r}%
^{\prime}\right)  }^{\left(  z\right)  }\left(  \overrightarrow{S}%
_{\overrightarrow{r}}\times\overrightarrow{S}_{\overrightarrow{r}^{\prime}%
}\right)  \cdot\widehat{z},\label{hdm}%
\end{equation}
where $\overrightarrow{S}_{\overrightarrow{r}}$ is a vector spin on site
$\overrightarrow{r}$ of a simple cubic crystal lattice, we assume positive
exchange parameters, $J_{0}$, $J>0$, the first sum is over nearest-neighbor
sites on the $x-y$ planes, and the second and third sums are over
nearest-neighbor sites along the $z$ direction. The last term in this spin
Hamiltonian provides the competition with the usual exchange term, which is
the essential ingredient to mimic a monoaxial DM model system.

We now make another simplifying assumption. Instead of considering continuous
variables, we assume discrete vector spin variables, with just six
possibilities, along the directions of the Cartesian axes. We then write the
six spin components,
\begin{equation}
\overrightarrow{S}_{\overrightarrow{r}}=%
\begin{pmatrix}
1\\
0\\
0
\end{pmatrix}
,\,%
\begin{pmatrix}
0\\
1\\
0
\end{pmatrix}
,\,%
\begin{pmatrix}
0\\
0\\
1
\end{pmatrix}
,\,%
\begin{pmatrix}
-1\\
0\\
0
\end{pmatrix}
,\,%
\begin{pmatrix}
0\\
-1\\
0
\end{pmatrix}
,\,%
\begin{pmatrix}
0\\
0\\
-1
\end{pmatrix}
.\label{S}%
\end{equation}
\newline Although this is a drastic simplification of the original spin model
system, it keeps the essential features of the DM interactions. This
formulation, however, is still not amenable to exact calculations. We then
resort to well-known mean-field approximate schemes, and analyze the main
features of the phase diagrams in terms of temperature and a parameter that
gauges the competition.

In Section 2, we formulate a solution of this problem on the basis of a
layer-by-layer mean-field calculation, which has been successfully used to
investigate the phase diagram of the ANNNI model \cite{Yokoi1981}. Although it
is simple to formulate and provides very sensible results, this mean-field
calculation demands a considerable numerical effort to unveil the fine details
of the phase diagram. We then restrict the analysis to the overall aspects of
the phase diagrams. In particular, we locate the paramagnetic transition
lines, which already indicate the presence of modulated structures.

In Section 3, we consider an analog of this model system on a Cayley tree. We
formulate the statistical problem as a non linear discrete map along the
generations of the tree. Attractors of this map, in the deep interior of the
tree, are known to correspond to physically realistic solutions, which come
from a standard pair approximation for the analogous model system on a Bravais
lattice. In the limit of infinite coordination of the tree, the problem
becomes considerably simpler, with solutions that should be close to the usual
mean-field results. According to earlier calculation for the analog of the
ANNNI model on a Cayley tree \cite{Yokoi1985}, we take advantage of this
infinite-coordination limit to analyze a number of fine details of the phase
diagram, including the onset of sequences of modulated structures and the
existence of devil's staircases of a fractal character.

\section{Layer-by-layer mean-field calculation}

We consider a cubic lattice of $N^{3}$ sites, and split the Hamiltonian
(\ref{hdm}) into two terms,%
\begin{equation}
\mathcal{H}=\mathcal{H}_{1}+\mathcal{H}_{2},
\end{equation}
where the first term includes the nearest-neighbor pair interactions on the
$x-y$ planes of the lattice,%
\begin{equation}
\mathcal{H}_{1}=-J_{0}\sum_{z=1}^{N}\sum_{\left(  x,y;\,\,x^{\prime}y^{\prime
}\right)  }\left[  S_{x,y,z}^{x}S_{x^{\prime},y\prime,z}^{x}+S_{x,y,z}%
^{y}S_{x^{\prime},y\prime,z}^{y}+S_{x,y,z}^{z}S_{x^{\prime},y\prime,z}%
^{z}\right]  ,
\end{equation}
and the second term refers to the interactions along the axial $z$ direction,%
\[
\mathcal{H}_{2}=-J\sum_{x,y}\sum_{z=1}^{N}\left[  S_{x,y,z}^{x}S_{x,y,z+1}%
^{x}+S_{x,y,z}^{y}S_{x,y,z+1}^{y}+S_{x,y,z}^{z}S_{x,y,z+1}^{z}\right]  -
\]%
\begin{equation}
-D\sum_{x,y}\sum_{z=1}^{N}\left[  S_{x,y,z}^{x}S_{x,y,z+1}^{y}-S_{x,y,z}%
^{y}S_{x,y,z+1}^{x}\right]  .
\end{equation}

We now use the Bogoliubov inequality,
\begin{equation}
G\left(  \mathcal{H}\right)  \leq G_{0}\left(  \mathcal{H}_{0}\right)
+\left\langle \mathcal{H}-\mathcal{H}_{0}\right\rangle _{0}=\Phi,
\end{equation}
where $G\left(  \mathcal{H}\right)  $ is the free energy associated with
$\mathcal{H}$, $G_{0}\left(  \mathcal{H}_{0}\right)  $ is the free energy
associated with a trial Hamiltonian $\mathcal{H}_{0}$, and $\left\langle
...\right\rangle _{0}$ is an expected value in a canonical ensemble defined by
the trial Hamiltonian. The mean-field approximation consists in assuming that
the free energy of this system comes from the minimization of the upper bound
$\Phi$ with respect to the parameters of the trial Hamiltonian.

The simplest form of a trial Hamiltonian, which still preserves the
possibility of modulated structures along the $z$ axis, is given by%
\begin{equation}
\mathcal{H}_{0}=-\sum_{x,y}\sum_{z}\overrightarrow{\eta}_{z}\cdot
\overrightarrow{S}_{x,y,z}=-\sum_{x,y}\sum_{z}\left[  \eta_{z}^{x}%
S_{x,y,z}^{x}+\eta_{z}^{y}S_{x,y,z}^{y}+\eta_{z}^{z}S_{x,y,z}^{z}\right]  ,
\end{equation}
where $\overrightarrow{\eta}_{z}=\left(  \eta_{z}^{x},\eta_{z}^{y},\eta
_{z}^{z}\right)  $ is a three-component trial field. We then have
\begin{equation}
G_{0}=-\frac{N^{2}}{\beta}\sum_{z}\ln\left[  2\cosh\left(  \beta\eta_{z}%
^{x}\right)  +2\cosh\left(  \beta\eta_{z}^{y}\right)  +2\cosh\left(  \beta
\eta_{z}^{z}\right)  \right]
\end{equation}
and
\[
\frac{1}{N^{2}}\left\langle \mathcal{H}-\mathcal{H}_{0}\right\rangle
_{0}=-2J_{0}\sum_{z}\left[  \left(  m_{z}^{x}\right)  ^{2}+\left(  m_{z}%
^{y}\right)  ^{2}+\left(  m_{z}^{z}\right)  ^{2}\right]  -
\]%
\[
-J\sum_{z}\left[  m_{z}^{x}m_{z+1}^{x}+m_{z}^{y}m_{z+1}^{y}+m_{z}^{z}%
m_{z+1}^{z}\right]  -
\]%
\begin{equation}
-D\sum_{z}\left[  m_{z}^{x}m_{z+1}^{y}-m_{z}^{y}m_{z+1}^{x}\right]  +\sum
_{z}\left[  \eta_{z}^{x}m_{z}^{x}+\eta_{z}^{y}m_{z}^{y}+\eta_{z}^{z}m_{z}%
^{z}\right]  ,
\end{equation}
in which we have introduced the definitions
\begin{equation}
m_{z}^{\nu}=\frac{\sinh\left(  \beta\eta_{z}^{\nu}\right)  }{\cosh\left(
\beta\eta_{z}^{x}\right)  +\cosh\left(  \beta\eta_{z}^{y}\right)
+\cosh\left(  \beta\eta_{z}^{z}\right)  },\label{mxdef}%
\end{equation}
where $\nu=x$, $y$, $z$, and $\beta=1/\left(  k_{B}T\right)  $ is the inverse
of temperature.

The minimization of $\Phi$ with respect to the trial fields $\left\{
\overrightarrow{\eta}_{z}\right\}  $ leads to the equations%
\begin{equation}
\eta_{z}^{x}=4J_{0}m_{z}^{x}+J\left(  m_{z-1}^{x}+m_{z+1}^{x}\right)
+D\left(  m_{z-1}^{y}-m_{z+1}^{y}\right)  ,\label{etax}%
\end{equation}%
\begin{equation}
\eta_{z}^{y}=4J_{0}m_{z}^{y}+J\left(  m_{z-1}^{y}+m_{z+1}^{y}\right)
-D\left(  m_{z-1}^{x}-m_{z+1}^{x}\right)  ,\label{etay}%
\end{equation}
and%
\begin{equation}
\eta_{z}^{z}=4J_{0}m_{z}^{z}+J\left(  m_{z-1}^{z}+m_{z+1}^{z}\right)
.\label{etaz}%
\end{equation}
We now insert these expressions for the trial fields into the definitions of
the local magnetizations, given by (\ref{mxdef}), and write an infinite system
of nonlinear coupled mean-field equations of state for the local magnetizations.

As in the mean-field calculations for the ANNNI model \cite{Yokoi1981}, given
the temperature and the parameters of the system, the problem consists in
finding a multiple set of numerical solutions, and choosing the solution that
minimizes the free energy. Except in the immediate neighborhood of a critical
line, this search for the physically acceptable solutions becomes a purely
numerical problem. Usually, we find a numerical solution with the assumption
of a periodicity of $L$ lattice spacings along the $z$ direction, and take
note of the associated value of the free energy. We then repeat this procedure
for a sequence of values of $L$. Physically acceptable solutions are shown to
minimize the free energy with respect to the length $L $. Instead of carrying
out this cumbersome numerical calculation, we now turn to an expansion of the
free energy.

It is easy to write the mean-field free energy as a power series in the spin
magnetizations, and make contact with a Landau-Ginzburg expansion. Keeping
terms up to order $4$, we have
\[
\frac{1}{N^{3}}\Phi=-\frac{1}{\beta}\ln6+\frac{1}{N}\sum_{z}\left(
-2J_{0}+\frac{3}{2\beta}\right)  \left[  \left(  m_{z}^{x}\right)
^{2}+\left(  m_{z}^{y}\right)  ^{2}+\left(  m_{z}^{z}\right)  ^{2}\right]  -
\]%
\[
-\frac{J}{N}\sum_{z}\left[  m_{z}^{x}m_{z+1}^{x}+m_{z}^{y}m_{z+1}^{y}%
+m_{z}^{z}m_{z+1}^{z}\right]  -\frac{D}{N}\sum_{z}\left[  m_{z}^{x}m_{z+1}%
^{y}-m_{z}^{y}m_{z+1}^{x}\right]  +
\]%
\begin{equation}
+\frac{9}{4\beta N}\sum_{z}\left[  \left(  m_{z}^{x}\right)  ^{2}\left(
m_{z}^{y}\right)  ^{2}+\left(  m_{z}^{x}\right)  ^{2}\left(  m_{z}^{z}\right)
^{2}+\left(  m_{z}^{y}\right)  ^{2}\left(  m_{z}^{z}\right)  ^{2}\right]  +...
\end{equation}
We then assume periodic boundary conditions, and use a Fourier representation,%
\begin{equation}
m_{z}^{\nu}=\frac{1}{\sqrt{N}}\sum_{q}m_{q}^{\nu}\exp\left(  iqz\right)  ,
\end{equation}
where $\nu=x,y,z$, and the sum is restricted to the first, and symmetric,
Brillouin zone. Also, it is convenient to write%
\begin{equation}
m_{q}^{\nu}=\frac{1}{\sqrt{2}}(R_{q}^{\nu}+iI_{q}^{\nu}),
\end{equation}
with%
\begin{equation}
R_{q}^{\nu}=R_{-q}^{\nu},\qquad I_{q}^{\nu}=-I_{-q}^{\nu}.
\end{equation}
Keeping terms up to second order, we have%
\[
\frac{1}{N^{3}}\Phi-\frac{1}{\beta}\ln6=
\]%
\[
=\frac{1}{2}\sum_{q}\left(  \frac{3}{2\beta}-2J_{0}-J\cos q\right)  \left[
\left(  R_{q}^{x}\right)  ^{2}+\left(  I_{q}^{x}\right)  ^{2}+\left(
R_{q}^{y}\right)  ^{2}+\left(  I_{q}^{y}\right)  ^{2}+\left(  R_{q}%
^{z}\right)  ^{2}+\left(  I_{q}^{z}\right)  ^{2}\right]  -
\]%
\begin{equation}
-D\sum_{q}\left(  \operatorname{sen}q\right)  \left[  I_{q}^{x}R_{q}^{y}%
-I_{q}^{y}R_{q}^{x}\right]  +\cdot\cdot\cdot,
\end{equation}
which is a quadratic form in terms of the real normal modes in Fourier space.
Note the symmetry along the $q$ axis. Also, note that the coupling is
restricted to the $x-y$ real modes only. It is then convenient to define
\begin{equation}
A=\frac{3}{4\beta}-J_{0}-\frac{1}{2}J\cos q;\qquad B=D\operatorname{sen}q,
\end{equation}
and to introduce the simplified notation%
\begin{equation}
x_{1}=R_{q}^{x};\quad x_{2}=I_{q}^{x};\quad y_{1}=R_{q}^{y};\quad y_{2}%
=I_{q}^{y}\quad z_{1}=R_{q}^{z};\quad z_{2}=I_{q}^{z}.
\end{equation}
We then have to analyze the quadratic form
\begin{equation}
Q=A\left(  x_{1}^{2}+x_{2}^{2}+y_{1}^{2}+y_{2}^{2}+z_{1}^{2}+z_{2}^{2}\right)
+B\left(  y_{2}x_{1}-x_{2}y_{1}\right)  =\left(  \overrightarrow{v}\right)
^{t}\mathbf{M}\left(  \overrightarrow{v}\right)  ,
\end{equation}
in which $\overrightarrow{v}$ is a column vector, $\widetilde{\mathbf{v}%
}=\left(  x_{1},x_{2},y_{1},y_{2},z_{1},z_{2}\right)  $, and the matrix
$\mathbf{M}$ is given by
\begin{equation}
\mathbf{M}=\left(
\begin{array}
[c]{cccccc}%
A & 0 & 0 & B/2 & 0 & 0\\
0 & A & -B/2 & 0 & 0 & 0\\
0 & -B/2 & A & 0 & 0 & 0\\
B/2 & 0 & 0 & A & 0 & 0\\
0 & 0 & 0 & 0 & A & 0\\
0 & 0 & 0 & 0 & 0 & A
\end{array}
\right)  .
\end{equation}
This matrix has two trivial real and degenerate eigenvalues,
\begin{equation}
\lambda_{1}=\lambda_{2}=A=\frac{3}{4\beta}-J_{0}-\frac{1}{2}J\cos q.
\end{equation}
The remaining eigenvalues are associated with a $4\times4$ supersymmetric
matrix. We then have%
\begin{equation}
\lambda_{3,4}=A\pm\frac{1}{2}B=\frac{3}{4\beta}-J_{0}-\frac{1}{2}J\cos
q\pm\frac{1}{2}D\sin q.
\end{equation}

Taking into account the form of these eigenvalues, it is immediate to obtain
an expression for the border of the disordered region,%
\begin{equation}
\frac{1}{\beta_{c}}=k_{B}T_{c}=\max\limits_{q}\left[  \frac{4}{3}J_{0}%
+\frac{2}{3}J\,\cos q\pm\frac{2}{3}D\,\sin q\right]  .
\end{equation}
For $D=0$, we recover the (mean-field) critical temperature of a simple ferromagnet,%

\begin{equation}
\frac{1}{\beta_{c}^{F}}=k_{B}T_{c}^{F}=\max\limits_{q}\left(  \frac{4}{3}%
J_{0}+\frac{2}{3}J\cos q\right)  =\frac{4}{3}J_{0}+\frac{2}{3}J,\label{tf}%
\end{equation}
For $D\neq0$, we have%
\begin{equation}
\tan q=\pm\frac{D}{J},
\end{equation}
which corresponds to propagating waves along the two directions of the $z$
axis, with the critical temperature%
\begin{equation}
\frac{1}{\beta_{c}^{M}}=k_{B}T_{c}^{M}=\frac{4}{3}J_{0}+\frac{2}{3}\left(
J^{2}+D^{2}\right)  ^{1/2}.\label{tm}%
\end{equation}

From equations (\ref{tf}) and (\ref{tm}) we have the paramagnetic transition
lines,%
\begin{equation}
k_{B}T_{c}^{F}-\frac{4}{3}J_{0}=\frac{2}{3}J\text{ },\qquad k_{B}T_{c}%
^{M}-\frac{4}{3}J_{0}=\frac{2}{3}\left(  J^{2}+D^{2}\right)  ^{1/2}.
\end{equation}
These equations indicate that there is always a transition to a spatially
modulated structure, except at a trivial, ferromagnetic, multicritical point,
at $D=0$. \medskip

\section{Analysis of the DM model on a Cayley tree}

We now turn to calculations for the analogous DM model on a Cayley tree, which
is perhaps the simplest way to unveil the rich structure of the ordered
phases. We then consider the monoaxial Hamiltonian,%
\begin{equation}
\mathcal{H}=-J\sum_{\left(  \overrightarrow{r},\overrightarrow{r}^{\prime
}\right)  }\overrightarrow{S}_{\overrightarrow{r}}\cdot\overrightarrow
{S}_{\overrightarrow{r}^{\prime}}-D\sum_{\left(  \overrightarrow
{r},\overrightarrow{r}^{\prime}\right)  }\left(  \overrightarrow
{S}_{\overrightarrow{r}}\times\overrightarrow{S}_{\overrightarrow{r}^{\prime}%
}\right)  \cdot\widehat{z},
\end{equation}
where $J>0$, and $\overrightarrow{r}$ and $\overrightarrow{r}^{\prime}$ are
nearest-neighbor sites along the successive generations of a Cayley tree (and
we discard ferromagnetic interactions between sites belonging to the same
generation of the tree). Although it is not entirely similar to the previously
analyzed mean-field model, the formulation on a Cayley tree is known to lead
to essentially the same qualitative results \cite{Oliveira1985}. Also, the
problem is written as a nonlinear dissipative discrete map, along the
generations of a Cayley tree, which is much easier to analyze than the
corresponding area-preserving map associated with the mean-field solutions.
Attractors of this dissipative mapping problem are known to mimic the phase
structure of the ordered region of the mean-field phase diagrams.%

%TCIMACRO{\FRAME{ftbpFU}{4.2627in}{2.6187in}{0pt}{\Qcb{Sketch of three
%generations of a Cayley tree with coordination three.}}{}{figure_cayley.png}%
%width 4.2627in;  height 2.6187in;  depth 0pt;  original-width 9.0728in;
%original-height 5.5521in;  cropleft "0";  croptop "1";  cropright "1";
%cropbottom "0";  filename 'figure_cayley.png';file-properties "XNPEU";}}}%
%BeginExpansion
\begin{figure}
[ptb]
\begin{center}
\includegraphics[
height=2.6187in,
width=4.2627in
]%
{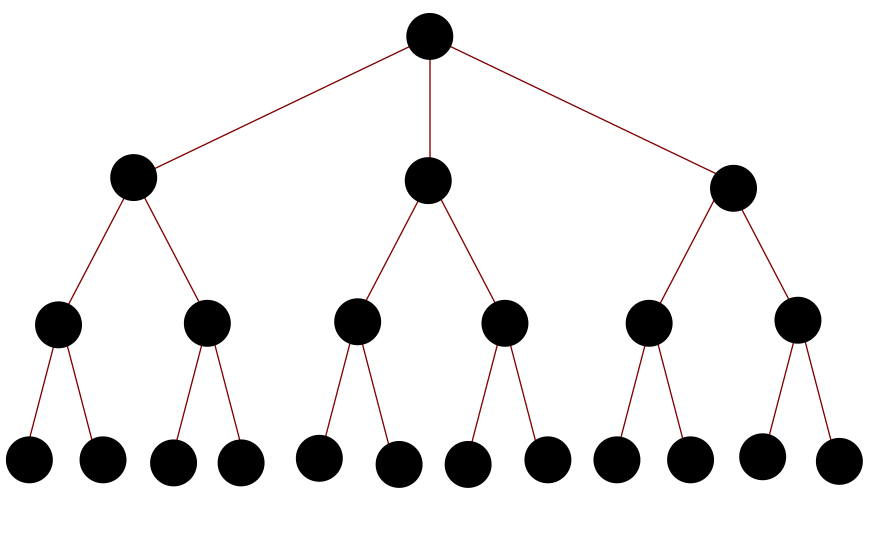}%
\caption{Sketch of three generations of a Cayley tree with coordination
three.}%
\end{center}
\end{figure}
%EndExpansion

In Figure 1, we sketch a Cayley tree of coordination $z=3$ (and ramification
$r=z-1=2$). We then consider a tree of arbitrary ramification $r$, and use
well-known procedures to write six recursion relations,
\begin{equation}
Z_{1}^{\prime}=\left(  e^{\beta J}Z_{1}+e^{\beta D}Z_{2}+Z_{3}+e^{-\beta
J}Z_{4}+e^{-\beta D}Z_{5}+Z_{6}\right)  ^{r},\label{Z1}%
\end{equation}%
\begin{equation}
Z_{2}^{\prime}=\left(  e^{-\beta D}Z_{1}+e^{\beta J}Z_{2}+Z_{3}+e^{\beta
D}Z_{4}+e^{-\beta J}Z_{5}+Z_{6}\right)  ^{r},\label{Z2}%
\end{equation}%
\begin{equation}
Z_{3}^{\prime}=\left(  Z_{1}+Z_{2}+e^{\beta J}Z_{3}+Z_{4}+Z_{5}+e^{-\beta
J}Z_{6}\right)  ^{r},\label{Z3}%
\end{equation}%
\begin{equation}
Z_{4}^{\prime}=\left(  e^{-\beta J}Z_{1}+e^{-\beta D}Z_{2}+Z_{3}+e^{\beta
J}Z_{4}+e^{\beta D}Z_{5}+Z_{6}\right)  ^{r},\label{Z4}%
\end{equation}%
\begin{equation}
Z_{5}^{\prime}=\left(  e^{\beta D}Z_{1}+e^{-\beta J}Z_{2}+Z_{3}+e^{-\beta
D}Z_{4}+e^{\beta J}Z_{5}+Z_{6}\right)  ^{r},\label{Z5}%
\end{equation}%
\begin{equation}
Z_{6}^{\prime}=\left(  Z_{1}+Z_{2}+e^{-\beta J}Z_{3}+Z_{4}+Z_{5}+e^{\beta
J}Z_{6}\right)  ^{r},\label{Z6}%
\end{equation}
where $Z_{i}^{\prime}$ is a partial partition function associated with a
central site in the spin state $i$, given by equation (\ref{S}), which is
related to the partial partition functions, $Z_{1}$ to $Z_{6}$, associated
with the nearest-neighbor sites belonging to the next generation of the tree.
The cycle-free structure of the Cayley tree enormously simplifies the form of
these recursion relations. It is worth to remark that, for $r=1$, we regain
the transfer matrix associated with the solution of the corresponding model on
a chain.

Although it is not difficult to analyze these recursion relations, we further
simplify the problem, and at the same time emphasize the analogy with the
mean-field results, by resorting to the infinite coordination limit of the
tree. We then take $r\rightarrow\infty$, and $J$, $D\rightarrow0$, but keep
fixed the products $Jr$ and $Dr$.

In this infinite-coordination limit, the problem is reduced to the analysis of
just three recursion relations, which involve the vector components of the
average of an \textquotedblleft effective magnetization\textquotedblright,
\begin{equation}
m_{x}^{\prime}=\frac{1}{M}\left[  \sinh(\beta Jrm_{x}+\beta Drm_{y})\right]
,\label{mx}%
\end{equation}%
\begin{equation}
m_{y}^{\prime}=\frac{1}{M}\left[  \sinh(-\beta Drm_{x}+\beta Jrm_{y})\right]
,\label{my}%
\end{equation}%
\begin{equation}
m_{z}^{\prime}=\frac{1}{M}\left[  \sinh(\beta Jrm_{z})\right]  ,\label{mz}%
\end{equation}
with the denominator
\begin{equation}
M=\cosh(\beta Jrm_{x}+\beta Drm_{y})+\cosh(-\beta Drm_{x}+\beta Jrm_{y}%
)+\cosh(\beta Jrm_{z}).\label{den}%
\end{equation}

We now use analytical and numerical techniques to analyze the attractors of
this map.

\subsection{Attractors of the map}

There is always a trivial solution, $m_{x}^{\ast}=m_{y}^{\ast}=m_{z}^{\ast}=0
$ of equations (\ref{mx})-(\ref{mz}). The linearization about this trivial
fixed point leads to the matrix equation%
\begin{equation}
\overrightarrow{m}^{\prime}=\mathbf{M}\,\overrightarrow{m},
\end{equation}
where%
\begin{equation}
M=\frac{1}{3}%
\begin{pmatrix}
\beta Jr & \beta Dr & 0\\
-\beta Dr & \beta Jr & 0\\
0 & 0 & \beta Jr
\end{pmatrix}
.
\end{equation}
Therefore, we have the eigenvalues
\begin{equation}
\lambda_{1}=\frac{1}{3}\beta Jr,\label{e1}%
\end{equation}%
\begin{equation}
\lambda_{2}=\frac{1}{3}(\beta Jr+i\beta Dr),\label{e2}%
\end{equation}
and%
\begin{equation}
\lambda_{3}=\frac{1}{3}(\beta Jr-i\beta Dr).\label{e3}%
\end{equation}
\newline The conditions of linear stability of this trivial (disordered) fixed
point are given by%
\begin{equation}
T>\frac{1}{3},\label{c1}%
\end{equation}
and
\begin{equation}
T>\frac{\left( 1+p^{2}\right) ^{1/2}}{3},\label{c2}%
\end{equation}
where we have introduced a simplified notation, $T=1/\left(  \beta Jr\right)
$ and $p=D/J$. We then have a similar picture as in the previous mean-field
calculation. There is always a paramagnetic transition, which is associated
with the complex eigenvalues of the trivial fixed point, except at $p=0$,
which leads to a simple ferromagnetic transition.

In order to go beyond the linear analysis, and investigate the stability of
additional attractors, we have to resort to numerical calculations. We then
characterized two types of behavior: (i) in one of these possibilities, the
magnetization settles in a specific value (one of the components settles in
the value $1$, and the other components vanish); (ii) the second possibility
is a cyclic pattern in the representation of $m_{x}$ versus $m_{y}$, as we
indicate in Figure 2 (the component $m_{z}$ of the magnetization flows to a
fixed value, as in the first case). The first possibility leads to a quite
trivial ordered structure. The second possibility, however, leads to a
spatially modulated structure.%

%TCIMACRO{\FRAME{ftbpFU}{4.2186in}{2.6195in}{0pt}{\Qcb{Graph of $200$
%iterations of $m_{y}$ versus $m_{x}$, for $T=0.2$ and $p=0.9$. We assume the
%initial values $m_{x}=m_{y}=m_{z}=1$, and discard the initial $10000$
%iterations. This modulated structure is characterized by the wave number of
%the main harmonic component, $q/2\pi=0.13$ (assuming a unit lattice
%paramater).}}{}{figure_modulated_rev.png}%
%width 4.2186in;  height 2.6195in;  depth 0pt;  original-width 12.9497in;
%original-height 8.009in;  cropleft "0";  croptop "1";  cropright "1";
%cropbottom "0";  filename 'figure_modulated_rev.png';file-properties "XNPEU";}%
%}}%
%BeginExpansion
\begin{figure}
[ptb]
\begin{center}
\includegraphics[
height=2.6195in,
width=4.2186in
]%
{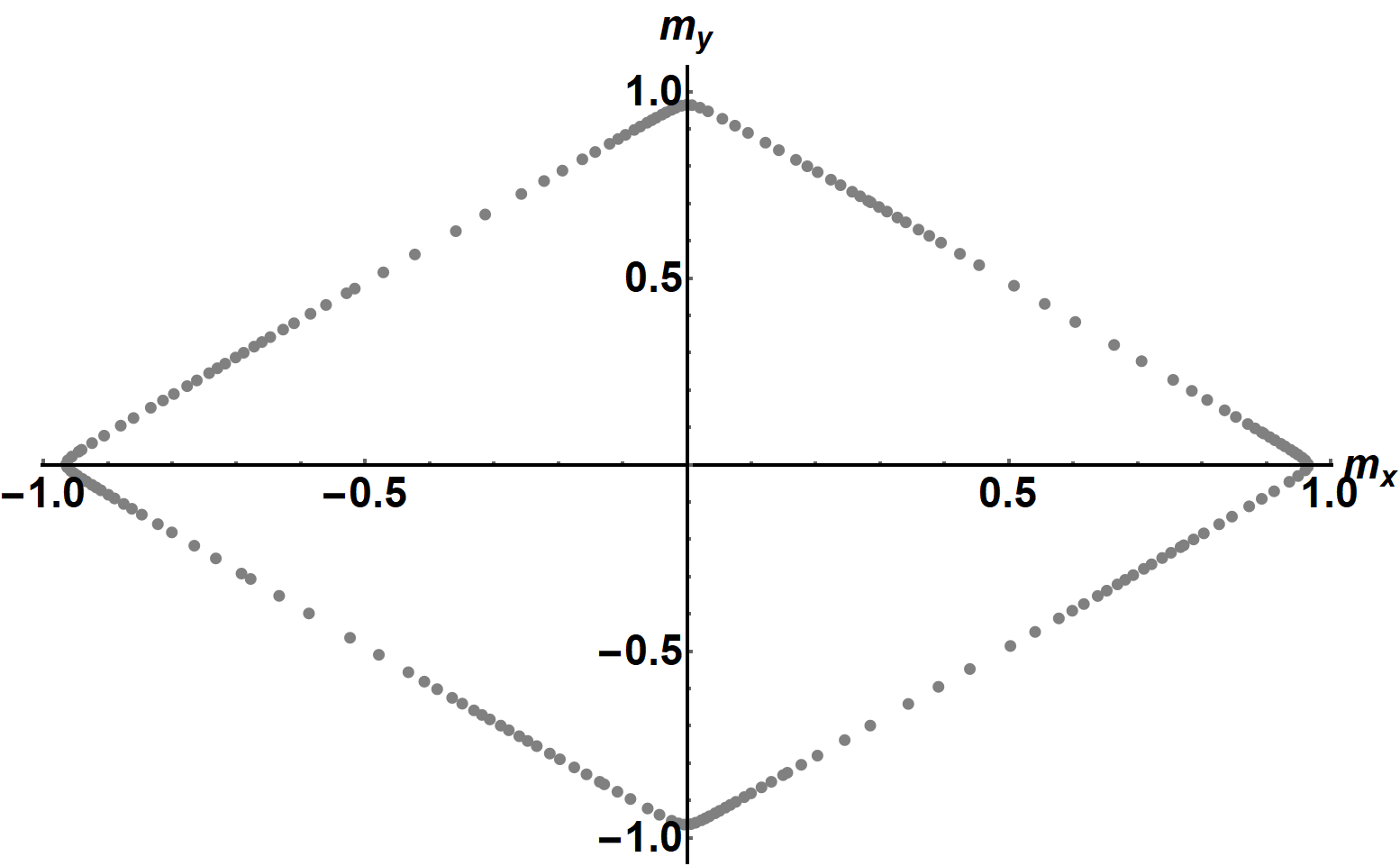}%
\caption{Graph of $200$ iterations of $m_{y}$ versus $m_{x}$, for $T=0.2$ and
$p=0.9$. We assume the initial values $m_{x}=m_{y}=m_{z}=1$, and discard the
initial $10000$ iterations. This modulated structure is characterized by the
wave number of the main harmonic component, $q/2\pi=0.13$ (assuming a unit
lattice paramater).}%
\end{center}
\end{figure}
%EndExpansion

We have used this numerical procedure to sketch the phase diagram of Figure 3.
Attractors of the map correspond to disordered (D), ferromagnetically ordered
(O), and spatially modulated (M) structures. In general, given the values of
reduced temperature $T$ and the parameter $p=D/J$, we choose arbitrary initial
conditions ($0<m_{x},m_{y},m_{z}<1$) and iterate equations (\ref{mx}) to
(\ref{mz}) about $10^{4}$ times. Final values of the magnetizations $m_{x}$
and $m_{y}$ are determined with a precision of about $10^{-5}$ (the component
$m_{z}$ does not display any modulation).%

%TCIMACRO{\FRAME{ftbpFU}{3.9686in}{2.6195in}{0pt}{\Qcb{Phase diagram in terms
%of temperature ($T$) and the parameter of chirality ($p=D/J$). We indicate
%disordered (D), ferromagnetically ordered (O) and some spatially modulated
%phases (in the white region). We use the dimensionless wave number, $q/2\pi$,
%which is written in terms of $2\pi$, to indicate some of the main modulated
%phases. In the groud state, for $p>1$, we have $q/2\pi=0.25$.}}{}%
%{novo_diagrama.png}{\special{ language "Scientific Word";  type "GRAPHIC";
%width 3.9686in;  height 2.6195in;  depth 0pt;  original-width 15.8226in;
%original-height 10.4063in;  cropleft "0";  croptop "1";  cropright "1";
%cropbottom "0";  filename 'Novo_diagrama.png';file-properties "XNPEU";}}}%
%BeginExpansion
\begin{figure}
[ptb]
\begin{center}
\includegraphics[
height=2.6195in,
width=3.9686in
]%
{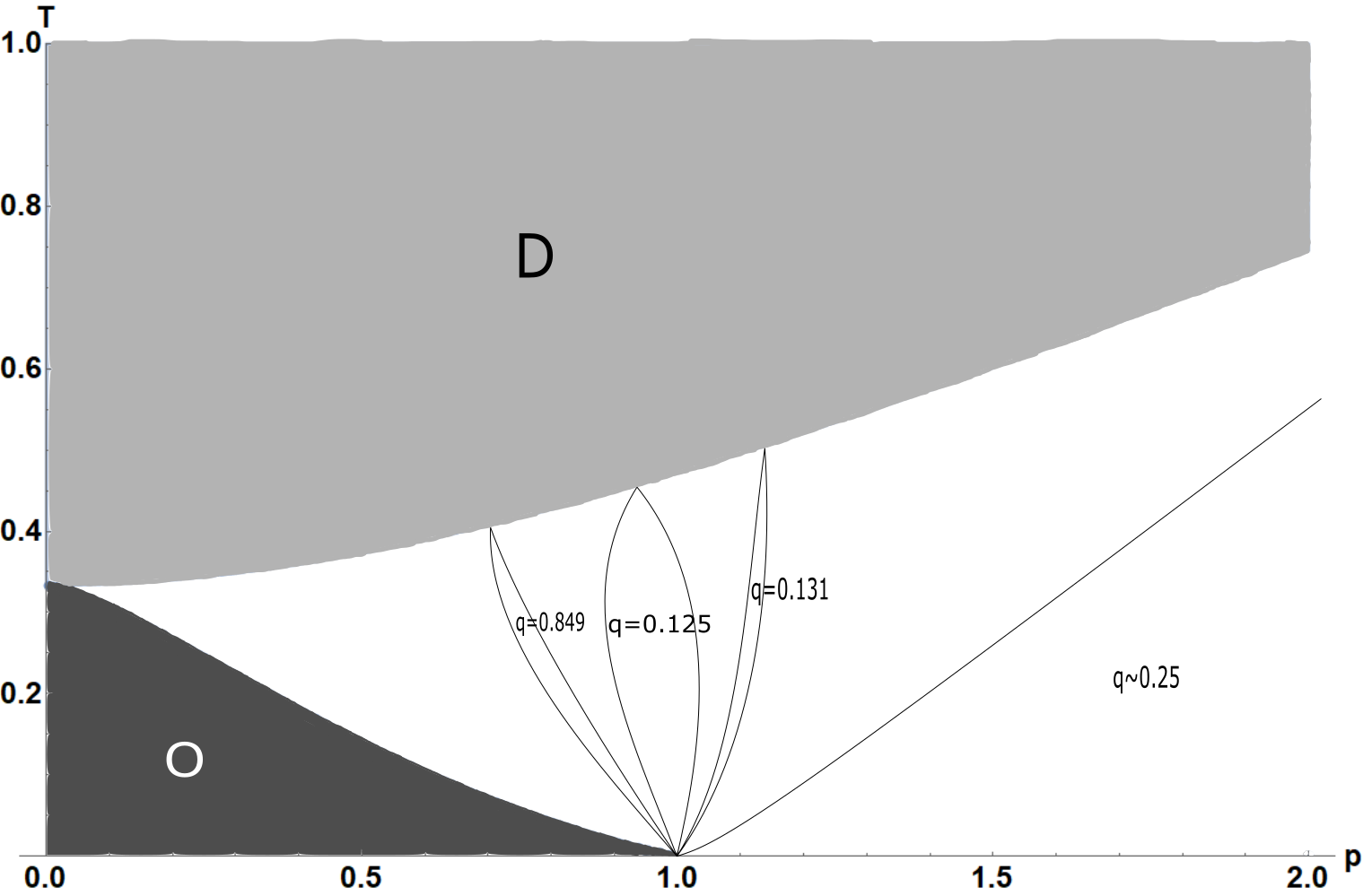}%
\caption{Phase diagram in terms of temperature ($T$) and the parameter of
chirality ($p=D/J$). We indicate disordered (D), ferromagnetically ordered (O)
and some spatially modulated phases (in the white region). We use the
dimensionless wave number, $q/2\pi$, which is written in terms of $2\pi$, to
indicate some of the main modulated phases. In the groud state, for $p>1$, we
have $q/2\pi=0.25$.}%
\end{center}
\end{figure}
%EndExpansion

\section{The devil's staircases}

The devil's staircase is a fractal structure that has been associated with the
sequences of modulated structures in the ordered region of the phase diagrams
of lattice statistical models, in particular the ANNNI model, with competing
interactions. We show that the same kind of behavior comes from the analysis
of the iterates of equations (\ref{mx}) to (\ref{mz}).

Given the parameters $T$ and $p$, we plot a graph of $m_{y}$ versus $m_{x}$,
as shown in Figure 2, and note that the attractor is a periodic function,
which gives rise to the definition of a period $q$ in units of $2\pi$. In a
more precise calculation, we perform a Fourier analysis of the attractor and
find the main Fourier component, which sets the period $q$. In Figure 4 we
plot the wave number $q$ versus the chiral parameter $p$, for a fixed value of
temperature, $T=0.2$, which already displays the characteristic shape of a
devil's staircase.%

%TCIMACRO{\FRAME{ftbpFU}{4.1009in}{2.6195in}{0pt}{\Qcb{Sketch of a devil's
%staircase at $T=0.2$.}}{}{figure_devil_staircase_rev.png}%
%width 4.1009in;  height 2.6195in;  depth 0pt;  original-width 13.7081in;
%original-height 8.7251in;  cropleft "0";  croptop "1";  cropright "1";
%filename 'figure_devil_staircase_rev.png';file-properties "XNPEU";}}}%
%BeginExpansion
\begin{figure}
[ptb]
\begin{center}
\includegraphics[
height=2.6195in,
width=4.1009in
]%
{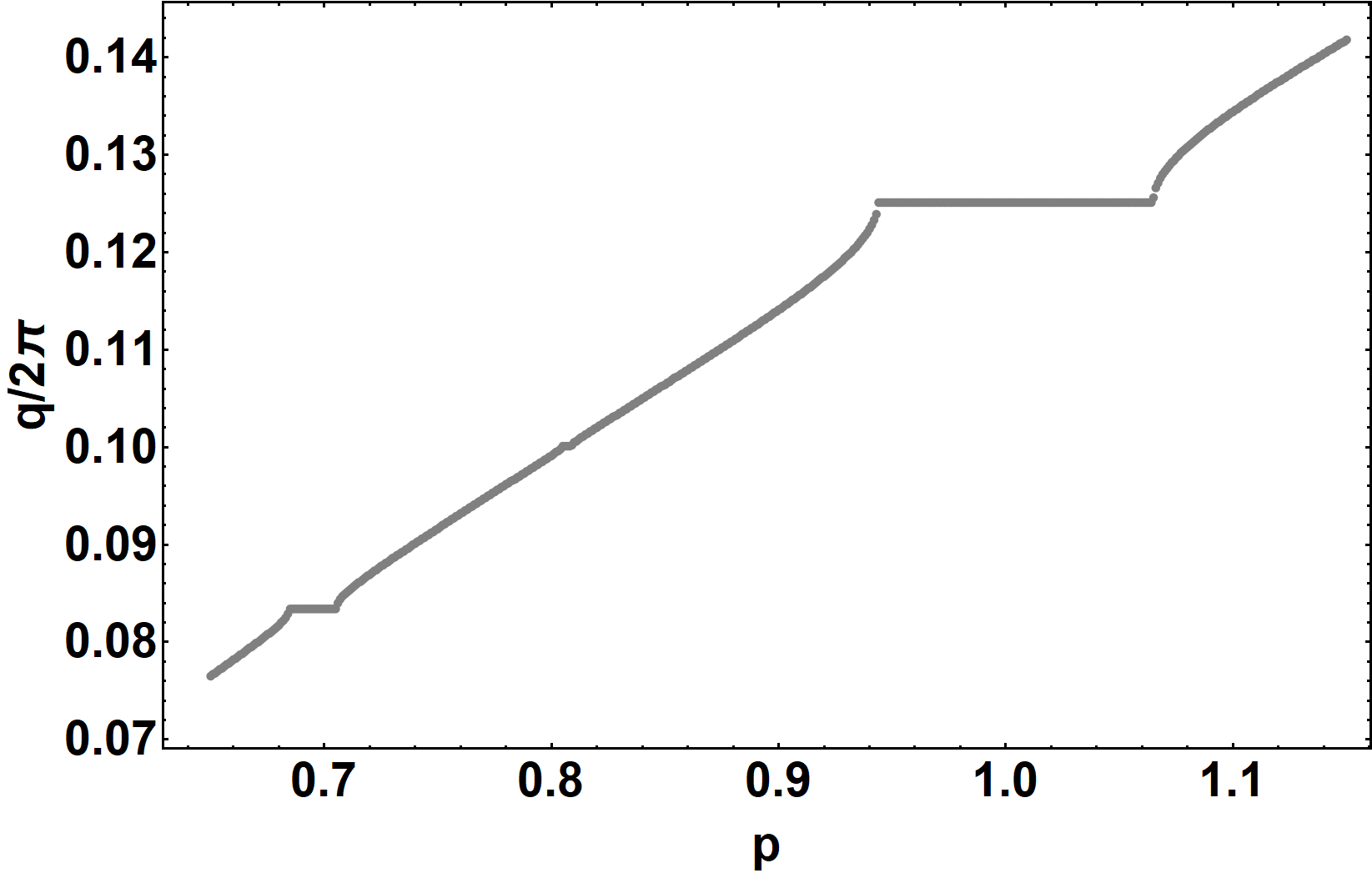}%
\caption{Sketch of a devil's staircase at $T=0.2$.}%
\end{center}
\end{figure}
%EndExpansion

We now turn to the calculation of the Hausdorff dimension associated with
these plots of $q$ versus $p$, which leads to the characterization of the
fractal nature of the devil's staircase. According to a box-counting algorithm
\cite{Nascimento2014}, we choose a value $\epsilon>0$, and then calculate the
sum of the step sizes, along the $p$ axis, of all steps with width (size along
the $p$ axis) larger than than $\epsilon$. We then subtract the total width of
the $p$ interval and obtain the function $X(\epsilon)$. The slope of a plot of
$\log(X(\epsilon)/\epsilon)$ versus $\log(1/\epsilon)$, in the limit of small
$\epsilon$, leads to an estimate of the Hausdorff dimension $D_{H}$ of the
staircase. In the modulated region, we found several examples of plots of $q$
versus $p$, at fixed temperature, with $D_{H}<1$, which characterizes a
fractal object. Also, we observed that $D_{H}$ increases with temperature,
which indicates a route to an incommensurate structure. In figure 5, we show
one of these plots of the Hausdorff dimension versus normalized temperature.%

%TCIMACRO{\FRAME{ftbpFU}{4.2376in}{2.6195in}{0pt}{\Qcb{Plot of the Hausdorff
%dimension as a function of normalized temperature $T$, calculated in the
%interval $0.65<p<1.15.$}}{}{figure_hausdorff_temp_rev.png}%
%width 4.2376in;  height 2.6195in;  depth 0pt;  original-width 15.4836in;
%original-height 9.5328in;  cropleft "0";  croptop "1";  cropright "1";
%filename 'figure_hausdorff_temp_rev.png';file-properties "XNPEU";}}}%
%BeginExpansion
\begin{figure}
[ptb]
\begin{center}
\includegraphics[
height=2.6195in,
width=4.2376in
]%
{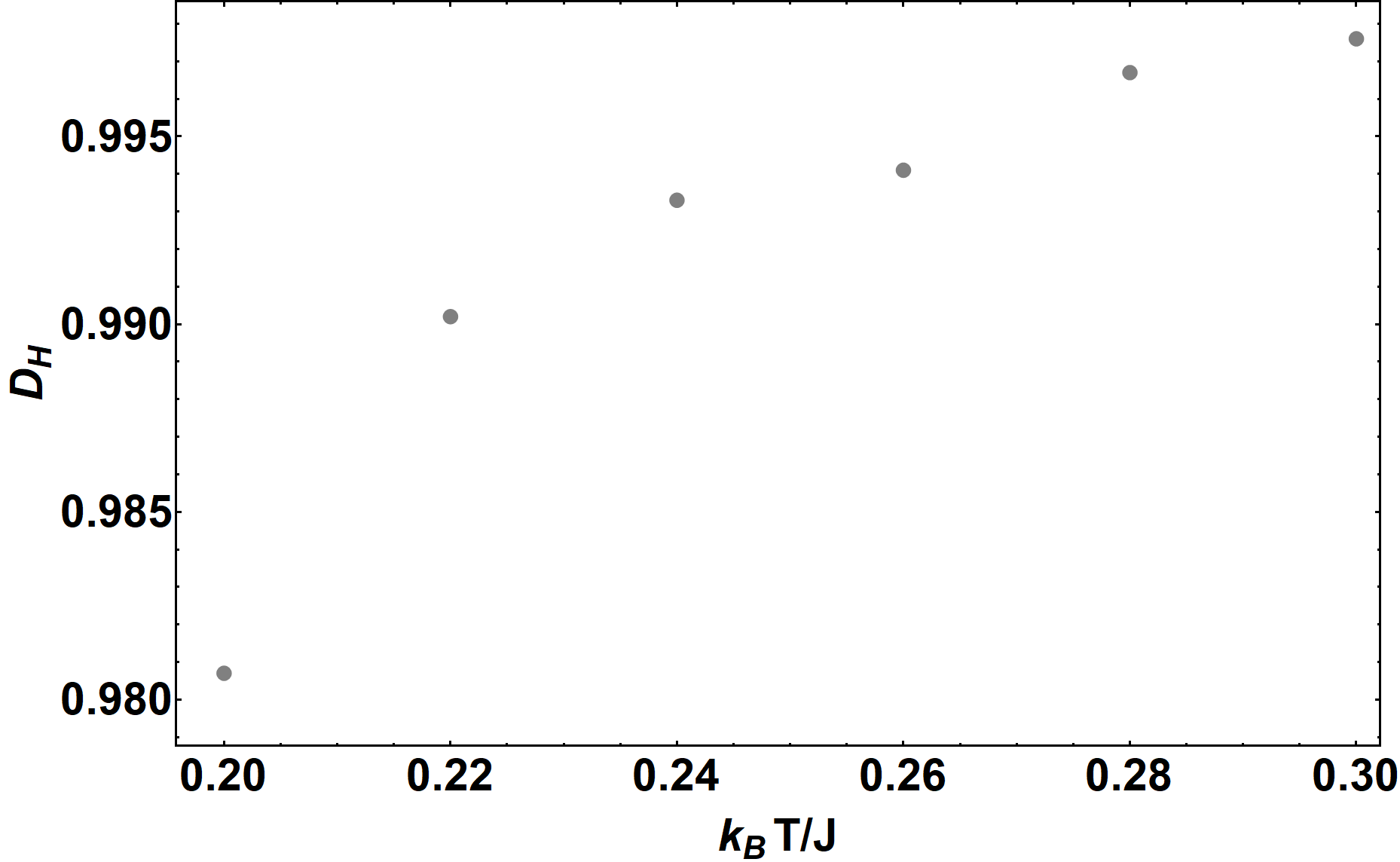}%
\caption{Plot of the Hausdorff dimension as a function of normalized
temperature $T$, calculated in the interval $0.65<p<1.15.$}%
\end{center}
\end{figure}
%EndExpansion

The increase of the Hausdorff dimension $D_{H}$ with $T$ ($D_{H}\rightarrow1 $
as $T\rightarrow T_{c}$) indicates the typical route to an incommensurate structure.

\section{Conclusions}

We used mean-field approximations to analyze the phase diagram of an
elementary system of discrete classical spin variables on a crystal lattice,
with ferromagnetic nearest-neighbor exchange interactions and the addition of
competing chiral interactions along an axis. This is perhaps the simplest
three-dimensional lattice system with the inclusion of Dzyaloshinskii-Moriya
interactions. We show that this monoaxial system displays the characteristic
modulated phases of helimagnetic compounds. Besides writing the equations of a
conventional mean-field layer-by-layer calculation, which already indicates
the existence of transitions to modulated structures, we performed some
detailed calculations for the analogous model on a Cayley tree, in the limit
of infinite coordination. Calculations on the tree confirm the general
features of the mean-field results, and provide a simpler way to check fine
details of the phase diagrams, in terms of temperature and a parameter of
chirality. In particular, we show the existence of many sequences of modulated
structures. Also, we confirm the existence of devil's staircases, which are
associated with a non-integer Hausdorff dimension.\bigskip

\textbf{Acknowledgment}

We acknowledge many fruitful discussions with Doctor Eduardo S. Nascimento. Also, we
acknowledge the financial support of the Brazilian agencies CNPq and CAPES.

\end{document}